\documentclass[12pt]{article}
\input epsf.sty
\def\ZZZ{{\hbox{ Z\kern-1.6mm Z}}}

\newcommand{\tl}{\wt\lambda}

\newcommand{\wt}{\widetilde}

\newcommand{\be}{\begin{equation}}
\newcommand{\ee}{\end{equation}}
\newcommand{\ben}{\begin{eqnarray}\displaystyle}
\newcommand{\een}{\end{eqnarray}}
\newcommand{\refb}[1]{(\ref{#1})}

\def\one{{\hbox{ 1\kern-.8mm l}}}
\def\zero{{\hbox{ 0\kern-1.5mm 0}}}
 
\begin{document}
{}~
{}~
\hfill\vbox{\hbox{hep-th/0308068}
}\break
 
\vskip .6cm
\centerline{\Large \bf 
Open-Closed Duality: Lessons from Matrix Model}

\vskip .6cm
\medskip

\vspace*{4.0ex}
 
\centerline{\large \rm
Ashoke Sen}
 
\vspace*{4.0ex}

\centerline{\large \it Harish-Chandra Research Institute}

\centerline{\large \it  Chhatnag Road, Jhusi,
Allahabad 211019, INDIA}
 
\centerline{E-mail: ashoke.sen@cern.ch,
sen@mri.ernet.in}
 
\vspace*{5.0ex}
 
\centerline{\bf Abstract} \bigskip

Recent investigations involving the decay of unstable D-branes in string
theory suggest that the tree level open string theory which describes the
dynamics of the D-brane already knows about the closed string states
produced in the decay of the brane. We propose a specific conjecture
involving quantum open string field theory to explain this classical
result, and show that the recent results in two dimensional string theory
are in exact accordance with this conjecture.

\vfill \eject
 
\baselineskip=18pt


Recent studies involving decay of unstable D-brane systems in string
theory indicate that while these D-branes are expected to decay into
closed string states of mass of order 
$1/g_s$\cite{0303139,0304192,0209222}, 
tree level
open 
string
theory provides an alternative description of the same 
process\cite{0203211,0203265,0204143,0207105,0208142,0202210,0209090,
0301038,0302146,0208196,0212248,0305177,0205085,0205098,0207107,
0304163,0301137,0301049,0303133}. 
In
particular various properties of the final state closed strings
produced during this decay agree with the predictions based on 
tree level open string analysis\cite{0304192,0305011,0306137,0306132}. 
These 
properties 
include the 
form 
of the
energy-momentum tensor, dilaton charge and anti-symmetric tensor field
charge of the system at late time. This suggests that in some way, 
tree level open string theory
already contains information about the final state
closed strings produced during this decay\cite{0305011,0306137}.

Clearly, in order to put this correspondence into a firmer footing, one 
needs a specific proposal for the full quantum theory. We propose the 
following 

\noindent {\it Conjecture: There is a quantum open string field theory
(OSFT) that describes the full dynamics of an unstable Dp-brane without an
explicit coupling to closed strings.  Furthermore, Ehrenfest theorem holds
in the weakly coupled OSFT: the classical results correctly
describe the evolution of the quantum expectation values.}

According to this conjecture, the effect of closed string emission is
already contained in the full quantum OSFT, and furthermore,
in the weak coupling limit, the results of quantum OSFT
must approach the results in classical OSFT.
Thus the above conjecture is sufficient to explain the observed open
closed duality mentioned earlier. For any finite 
coupling, the Ehrenfest theorem could break down over a sufficiently long 
time scale, but this time scale should approach infinity in the limit of 
zero coupling constant.

It is instructive to apply the above conjecture to the specific case of 
unstable D0-brane system. The quantum OSFT on a 
D0-brane is a quantum mechanical system of infinitely many degrees of 
freedom. 
If the above conjecture is valid, then this quantum mechanical system 
contains complete description of the closed strings produced in
the decay of the D0-brane, even though these closed strings 
live in the full space-time of string theory, which is (25+1) 
dimensional for the bosonic string theory and (9+1)-dimensional for the 
superstring theory. 

Note that the conjecture stated above does not imply that the OSFT
on a D0-brane (or any D$p$-brane for that matter) contains 
complete information about {\it all states} in string theory. It simply 
states 
that the 
OSFT
on the D0-brane is a consistent quantum theory by 
itself, and hence has included in it a description of the closed string 
states into which an unstable D0-brane is allowed to decay. In other 
words, OSFT on an unstable D-brane describes a closed subsector of the 
full string theory.

At present in the critical string theory there is not much further 
evidence for this conjecture beyond those already mentioned. The only 
other piece of information which is relevant is that formally the 
perturbation expansion of the OSFT around the 
maximum 
of the tachyon potential seems to be complete, in the sense that it 
reproduces correctly the Polyakov amplitudes of the first quantized string 
theory involving external open string states to all orders in perturbation 
theory.\footnote{This has been established\cite{GIDMARWIT,ZWIE91} for the 
cubic 
bosonic 
OSFT proposed by Witten\cite{WITTENSFT}, and is expected to 
hold\cite{9912120} 
also for 
the 
open 
superstring field theory proposed by 
Berkovits\cite{9503099,0001084,0002211}.} 
In particular the amplitude 
has the correct poles 
corresponding to intermediate closed string states\cite{thorn}. This 
suggests that the 
quantum OSFT is a consistent quantum theory by 
itself. 
Note however that the perturbation expansion discussed above is 
purely 
formal, as the amplitudes are divergent due to the presence of the 
open string tachyonic mode which lives on the unstable D-brane. 
Nevertheless, the formal consistency of the 
perturbation 
theory suggests that the same theory, when quantized correctly by 
expanding the action around the tachyon 
vacuum, will give a fully consistent quantum theory, as the tachyonic mode 
will be absent around such a vacuum.

We can however do much better in the two dimensional string theory for 
which a specific non-perturbative formulation is available in the form of 
a matrix quantum mechanics. There are two specific 
models for which 
the correspondence has been established, -- the two dimensional bosonic 
string theory\cite{GROMIL,BKZ,GINZIN} and the two dimensional type 0B 
string theory\cite{0307083,0307195}. 
Since 
the 
matrix models associated with the two systems are very similar, our 
discussion will be valid for both theories. 
In order that the various formul\ae\ in the two theories look identical, 
we 
shall set $\hbar=c=1$ and choose $\alpha'=1$ unit for the bosonic string 
theory and 
$\alpha'=1/2$ unit for type 0B string theory. In this convention the open 
string tachyon on the D0-brane has mass$^2=-1$ in both theories.
Also we shall define the closed string coupling constant $g_s$ in such 
a way that the D0-brane has mass $1/g_s$ in both 
theories.\footnote{There is an unresolved issue here. To the best of our 
knowledge it 
has not been shown that the D0-brane mass computed in the continuum 
string theory (which could be defined {\it e.g.} as the height of 
the maximum of the open string tachyon potential above the tachyon vacuum) 
agrees exactly with the prediction from 
the matrix model (the height of the maximum of the tachyon potential above 
the fermi level). Since the continuum string coupling constant is known in 
terms of the height of the tachyon potential in the matrix 
model(see {\it e.g.} \cite{9108019}),
this
problem could in principle be solved. We shall proceed by assuming that 
the D0-brane mass computed in the continuum string theory agrees with the 
corresponding answer in the matrix model.} We shall 
restrict our discussion mainly to D0-branes in type 0B string theory, 
since the two dimensional bosonic string theory is believed to be 
non-perturbatively inconsistent\cite{MOORE,9111035,9411028,9507041}.

\begin{figure}[!ht]
\leavevmode
\begin{center}
\epsfysize=5cm
\epsfbox{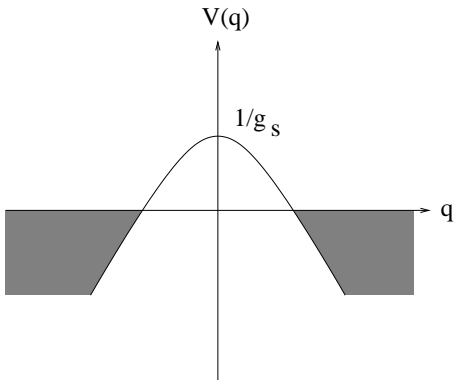}
\end{center}
\caption{The matrix model description of the vacuum state of type 0B 
string theory in 1+1 dimensions. All the negative energy states are filled 
as shown by the shaded region of the diagram.
} \label{f0}
\end{figure}
\begin{figure}[!ht]
\leavevmode
\begin{center}
\epsfysize=5cm
\epsfbox{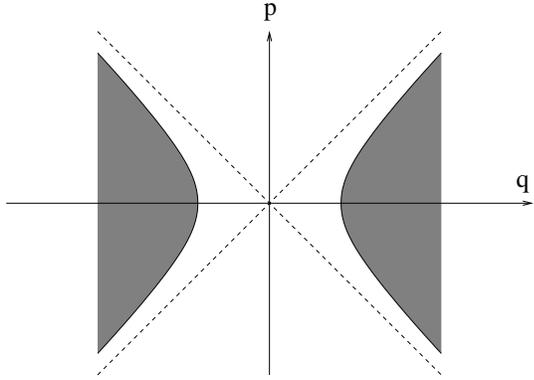}
\end{center}
\caption{Semi-classical representation of the vacuum state in the matrix 
model.} \label{f1}
\end{figure}

According to the matrix model - 
string theory correspondence, the two dimensional type 0B string theory is 
equivalent to a theory of infinite number of non-interacting fermions, 
each moving in an 
inverted harmonic oscillator potential with hamiltonian
\be \label{e1}
h(p,q) = {1\over 2} (p^2 - q^2) + {1\over g_s}\, ,
\ee
where $(q,p)$ denote a
canonically conjugate pair of variables.  The coordinate 
variable $q$ is related to the eigenvalue of an infinite dimensional 
matrix, but this information will not be necessary for our discussion. 
Clearly $h(p,q)$ has a continuous energy spectrum spanning the range 
$(-\infty, \infty)$.
The 
vacuum of the theory corresponds to all states with negative $h$ 
eigenvalue 
being filled and all states with positive $h$ eigenvalue being empty (see 
Fig.\ref{f0}). Thus 
the fermi surface is the surface of zero energy. In the semi-classical 
limit, in which we represent a quantum state by an area element of size 
$\hbar$ in the phase space spanned by $p$ and $q$, we can represent the 
vacuum by having the region $(p^2 - q^2) \le -{2\over g_s}$ 
filled, and rest of the region empty\cite{POLCH,9212027}. This has been 
shown in 
Fig.\ref{f1}. Thus in this picture the fermi 
surface in the phase space 
corresponds to the curve:\footnote{
The matrix model description for the two dimensional bosonic string theory 
is almost 
identical, except that only the 
$q<0$, $(p^2 - q^2) \le -{2\over g_s}$ region is filled, but the 
$q>0$
region is not filled. Clearly such a configuration is non-perturbatively 
unstable since the fermions on the left side of the potential could tunnel 
to the right side. For this reason the bosonic string theory in two 
dimensions is thought to be non-perturbatively inconsistent.
}
\be \label{e2}
{1\over 2} (p^2 - q^2) + {1\over g_s} = 0\, .
\ee

It has been realized 
recently\cite{0305148,0306177,0304224,0305194,0305159,0307083,0307195} 
that 
D0-branes 
in two dimensional 
string theory\cite{0101152,0202032,0202043} also have simple 
description in the matrix model. 
In particular, a
state of a single D0-brane of the theory 
corresponds to a single fermion excited from the fermi surface to some 
energy above zero. Since the fermions are non-interacting, these states do 
not mix with any other states in the theory (say with states where two or 
more fermions are excited above the fermi level or states where a fermion 
is excited from below the fermi level to the fermi level). As a result, 
the 
quantum states of a D0-brane are in one to one correspondence with the 
quantum states of the Hamiltonian $h(p,q)$ given in \refb{e1} with one 
crucial difference, -- the spectrum is cut off sharply for energy below 
zero due to Pauli exclusion principle.
Thus in the matrix model description, the quantum `open string field 
theory'
for a single 
D0-brane is described by the 
inverted harmonic oscillator hamiltonian \refb{e1} with all the negative 
energy states removed by hand. The 
classical limit of this quantum 
Hamiltonian is described by the classical Hamiltonian \refb{e1}, with a 
sharp 
cut-off on the phase space variables:
\be \label{e4}
{1\over 2} (p^2 - q^2) + {1\over g_s} \ge 0\, .
\ee
This is the matrix model description of classical `open string field 
theory' describing the dynamics of
a D0-brane.
\begin{figure}[!ht]
\leavevmode
\begin{center}
\epsfysize=5cm
\epsfbox{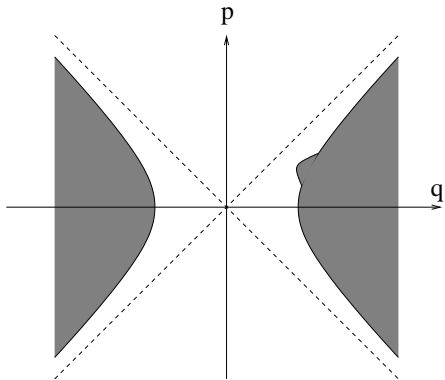}
\end{center}
\caption{Semiclassical representation of a closed string field 
configuration in the matrix model.} \label{f2}
\end{figure}

\begin{figure}[!ht]
\leavevmode
\begin{center}
\epsfysize=5cm
\epsfbox{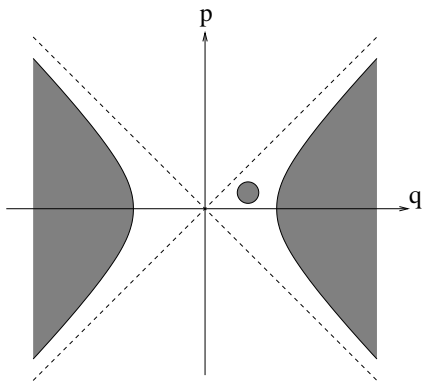}
\end{center}
\caption{Semi-classical representation of a state of the D0-brane in the 
matrix model.} \label{f3}
\end{figure}

Clearly the quantum system described above provides us with a 
complete description of
the dynamics of a single D0-brane. This is in accordance with 
the general conjecture put forward at the beginning of this note. 
Note in particular that there is no need to couple this system explicitly 
to closed strings. In this context 
we note that 
according to \cite{POLCH,9212027}, classical closed string field 
configurtions in 
this 
theory correspond to deformations of the fermi surface \refb{e2} in the 
phase space (see Fig.\ref{f2}). In contrast the semi-classical description 
of a 
D0-brane with energy of order $1/g_s$ amounts to filling up an area of
order $\hbar$ in the phase space at an energy of order $1/g_s$ above the 
fermi
surface (see Fig.\ref{f3}), and hence such a state cannot be described as 
a deformation of 
the fermi surface. Thus in general a D0-brane cannot be described as a 
classical closed string field configuration. However in the asymptotic 
past and asymptotic future all trajectories in the phase space, including 
the fermi surface, approach 
the asymptotes $p=\pm q$. Thus in this limit the D0-brane can be thought 
of as a deformation of the fermi surface, {\it i.e.} a classical closed 
string field configuration\cite{9212027}. The precise form of this closed 
string field 
configuration can be found using the bosonization 
formula\cite{DASJEV,SENWAD,GROSSKLEB} and was 
shown to agree\cite{0305159,0307083,0307195,0305194,0308047} with the 
coherent 
closed string 
fields produced in 
the decay of the D0-brane computed directly from the continuum string 
theory\cite{0303139,0304192}. 
This is one of the compelling pieces of evidence that the identification 
of the D0-brane with the single excited fermion in the matrix model 
description is correct.
But this also clearly demonstrates that closed strings produced in the 
`decay' of the D0-brane are already included in the quantum `open 
string field theory'
describing the dynamics of the D0-brane, and there is no 
need to take into account the closed string emission effect separately.

It remains to see how this simple system described by 
the Hamiltonian \refb{e1} with the phase space cut-off \refb{e4} is 
related to the more conventional description of the continuum open string 
field 
theories (OSFT)
of the type described in \cite{WITTENSFT} and 
\cite{9503099,0001084,0002211}.\footnote{I wish to thank L. Rastelli for 
pointing out that OSFT on a D0-brane in non-critical string theories can 
be formulated in the same way as in the case of critical string theories.
The essential point to note is that while the presence of the linear 
dilaton background changes the structure of the inner product in the 
closed string sector (so that the BPZ inner product between the SL(2,C) 
invariant vacuum states vanish), the structure of inner products in the 
open string sector on a D0-brane remains unchanged since these open 
strings do 
not carry any momentum in the Liouville direction.} To this end we note 
the following facts: \begin{enumerate}
\item The effective Planck's constant (coupling constant) of the single 
fermion quantum 
mechanics described in \refb{e1} is of order $g_s$\cite{POLCH}. This can 
be 
seen by introducing new variables $\wt p=\sqrt{g_s}p$, $\wt q = \sqrt{g_s} 
q$
so that the Hamiltonian expressed in terms of $\wt p$, $\wt q$ 
has an overall multiplicative factor of $1/g_s$, but otherwise there is no 
$g_s$ 
dependence either in the Hamiltonian or in the constraint \refb{e4}. 
The commutator of $\wt q$ and $\wt p$ is 
proportional to $i g_s$, and hence in terms of the rescaled variables 
$g_s$ 
is the effective Planck's constant. On the other hand in the standard 
OSFT
also $g_s$ is the effective Planck's constant since 
$1/g_s$ appears as an overall multiplicative factor in the action. 
Thus we see that $g_s$ plays the 
same role in the quantum 
OSFT
and the quantum system described by \refb{e1}, \refb{e4}.

\item Both the classical Hamiltonian \refb{e1} with the constraint 
\refb{e4} and the OSFT on a D0-brane have a lower bound of zero on the 
energy\cite{0001084,0002211,0003220,0004015,0010108,
0012198,0012210,0106231}.\footnote{For 
bosonic string 
theory, in both descriptions 
there is a local minimum at the zero of the energy, but globally the 
energy is unbounded from 
below\cite{KS,9605088,9912249,0002237,0208149,0211012,0009103,0009148,
0009191}.}

\item Next we shall compare time dependent classical solutions in the OSFT
on a D0-brane and the system described by \refb{e1}, 
\refb{e4}. During this discussion we shall assume that given any 
boundary conformal field theory (BCFT) obtained by deforming the original 
D0-brane BCFT by a marginal deformation, we have a classical solution of 
the OSFT. We should caution the reader however that explicit construction 
of these solutions may involve 
subtle issues, and so far a clear correpondence 
between BCFT and the classical solutions of OSFT 
have not been established in the context of time 
dependent solutions\cite{0207107,0304163}.

Comparing the classical solutions of the equations of motion derived from 
the Hamiltonian \refb{e1} subject to the constraint \refb{e4}, and the 
BCFT's describing time dependent configurations on a D0-brane, we find 
that
both systems have a continuous family of solutions labelled by 
the energy $E$ for all $E\ge 0$. In fact for each energy there are two 
inequivalent orbits. In the case of the system described by 
\refb{e1}, \refb{e4}, these solutions are given by:
\ben \label{e5}
&&q = \pm\sqrt{2(g_s^{-1} - E)} \, \cosh(x^0), \quad 
p = \pm \sqrt{2(g_s^{-1} - E)} \, \sinh(x^0), \quad \hbox{for 
$0\le E\le g_s^{-1}$}\, 
, \nonumber \\
&&q = \pm\sqrt{2(E-g_s^{-1})} \, \sinh(x^0), \quad p = 
\pm\sqrt{2(E-g_s^{-1})} 
\, 
\cosh(x^0),\quad \hbox{for $E\ge g_s^{-1}$} \, . \nonumber \\
\een
Here $x^0$ is the time coordinate.
On the other hand in OSFT for D0-branes in the continuum type 0B string 
theory, these solutions 
correspond 
to adding to the world-sheet theory a boundary deformation
proportional 
to\cite{0203265}\footnote{For bosonic string theory, the 
corresponding boundary 
deformations take the form\cite{0203211}:
\ben 
&&\pm \tl \cosh (X^0), \qquad \tl=\cos^{-1}(\sqrt{Eg_s}), \quad 
0\le\tl\le{1\over 2}, \qquad \hbox{for 
$0\le E\le g_s^{-1}$}, \nonumber \\
&&\pm \tl\sinh(X^0), \qquad \tl=\cosh^{-1}(\sqrt{Eg_s}), \quad \tl\ge 0,
\qquad \hbox{for 
$E\ge g_s^{-1}$} \, . \nonumber 
\een
} 
\ben \label{e6a}
&&\pm \tl \Psi^0 \, \sinh (X^0)\otimes \sigma_1, \quad 
\tl=\cos^{-1}(\sqrt{Eg_s}), \quad 
0\le\tl\le{1\over 2},  \quad 
\hbox{for
$0\le E\le g_s^{-1}$}\, , \nonumber \\
&&\pm \tl \Psi^0 \cosh(X^0)\otimes \sigma_1, \quad 
\tl=\cosh^{-1}(\sqrt{Eg_s}), \quad 
\tl\ge 0, \quad 
\hbox{for
$E\ge g_s^{-1}$} \, .
\een
Here
$X^0$ is the world-sheet field 
corresponding to the time coordinate, $\Psi^0$ denotes the world-sheet 
superpartner of $X^0$ and 
$\sigma_1$ is a Chan-Paton factor. 
In both the matrix model description based on \refb{e1}, \refb{e4} and the 
continuum string theory description, each of these orbits are open 
orbits, {\it i.e.}
they are not periodic. Thus the two theories have exactly the same family 
of classical solutions.

If classical OSFT
could be viewed as a Hamiltonian 
system described by a pair of phase space coordinates $(T,\Pi)$ with some 
Hamioltonian $H(\Pi, T)$ ({\it e.g.} $H=\sqrt{\Pi^2 + (V(T))^2}$ with 
$V(T)=1/(g_s \cosh T)$ as described in 
\cite{0204143,9909062,0003122,0003221,0004106,0009061,0209122,0301076,
0303035,0304045,0209034,0305229,0208019})
then the 
above correspondence between classical solutions would immediately imply 
that there is a canonical transformation relating this system to the one 
described by \refb{e1}, \refb{e4}. This canonical transformation can be 
found as follows. Let the trajectories of OSFT for a given energy $E$ be 
given by 
\be \label{e7}
T = F(E, x^0), \qquad \Pi = G(E, x^0)\, ,
\ee
and the trajectories of the system described by \refb{e1}, \refb{e4} be 
given by:
\be \label{e8}
q = f(E, x^0), \qquad p = g(E, x^0)\, .
\ee
We can now eliminate $E$ and $x^0$ from eqs.\refb{e7} and \refb{e8} to 
express $q$ and $p$ in terms of $T$ and $\Pi$ or vice versa. As long as 
the orbits are open, this is always possible and gives a one to one 
mapping 
between the allowed regions of the $(q,p)$ plane and the $(T,\Pi)$ 
plane. In particular, if the $(T,\Pi)$ coordinates are unconstrained, -- 
as in the case of the tachyon `effective Hamiltonian' $H=\sqrt{\Pi^2 + 1/ 
(g_s^2 \cosh^2 T)}$, -- then the region \refb{e4} will be mapped to the 
full $(T,\Pi)$ plane. The transformation 
constructed this way is also guaranteed to be canonical, and maps 
$H(\Pi,T)$ to $h(p,q)$. Thus the two 
systems are related by canonical transformation. (In contrast if the 
orbits had been periodic, such a transformation can be constructed only if 
the periods of the orbit for any given energy $E$ are 
identical 
in the two systems.)

Unfortunately OSFT in its current form cannot be thought of as a
Hamiltonian system since it has interaction terms involving higher 
order time derivatives.
Nevertheless, the fact that the classical solutions in the two systems are
in one to one correspondence strongly suggests that the two systems are
classically equivalent.

\item We can also compare the set of classical solutions in the euclidean 
version of the two theories. While the euclidean solutions are not 
directly relevant for comparing the two classical theories in the 
Lorentzian signature space-time, in the quantum theory these euclidean 
solutions induce tunnelling between two sides of the tachyon potential for 
orbits with $E<g_s^{-1}$, and hence comparing them between the two 
theories is important for establishing the equivalence between the two 
quantum theories\cite{0305159,0307195}. Euclideanization of the system 
described by \refb{e1}, 
\refb{e4} is achieved by making the replacement $p\to ip$, $x^0\to ix$ in 
the classical solutions and the constraint \refb{e4}. Thus the 
inequivalent classical 
solutions in this theory are:
\be \label{e11}
q = \lambda \cos x, \qquad p = \lambda\sin x, \qquad \lambda^2<{2\over 
g_s}\, .
\ee
On the other hand in the continuum string theory the euclidean solutions 
are obtained as boundary deformation of the world-sheet theory with $X^0$ 
and $\Psi^0$ replaced by $iX$ and $i\Psi$ respectively. The inequivalent 
classical solutions on a D0-brane in continuum type 0B string theory 
correspond to deformation by\cite{9808141,9812031}\footnote{
For bosonic string theory the corresponding boundary operator is $\tl \cos 
X$\cite{9402113,9404008,9811237,9902105}.
}
\be \label{e12}
\tl \Psi \sin X \otimes \sigma_1, 
\qquad 0\le \tl\le {1\over 2}\, .
\ee
The important point to note is that in both the matrix theory version and 
the continuum version the solutions are periodic in euclidean time 
coordinate $x$ with the same (energy independent) periodicity $2\pi$. Thus 
even in the euclidean 
theory the classical solutions of the system described by \refb{e1}, 
\refb{e4}
are in one to one correspondence to the classical solutions of OSFT. Had 
OSFT been described by a canonical Hamiltonian this would imply that the 
semiclassical tunnelling probability $P(E)$ for tunnelling across the 
potential barrier
at any given energy $E$ is identical in OSFT and 
the matrix model, -- $P(E)\sim \exp[-2\pi({1\over g_s}-E)]$, -- since for 
a canonical system $P(E)$ is determined in 
terms of the period $\tau(E)$ of the euclidean orbit via the relation 
$\tau(E)={d\over d E} \ln P(E)$. (In this context note that the tachyon 
`effective Hamiltonian' $H=\sqrt{\Pi^2 + 1 / (g_s^2 \cosh^2T)}$ also has 
orbits of the same period in the Euclidean theory\cite{0303139}. Hence 
the semiclassical 
tunnelling probability across the potential barrier in this theory is 
identical to that for the system described by \refb{e1}, \refb{e4}, {\it 
i.e.} $P(E)\sim \exp[-2\pi({1\over g_s}-E)]$.)

There is however one subtle issue here. The solution associated with
$\lambda=\sqrt{2/g_s}$ in the matrix model description corresponds to the
point $\tl={1\over 2}$ in the continuum OSFT description. Physically in
the continuum theory this solution represents a periodic array of 
D-instanton - anti-D-instanton pair
with periodicity $2\pi$. But given this configuration, we can deform it to
construct other solutions of the Euclidean OSFT where the array has a
different periodicity. In fact we can construct a family of solutions
parametrized by the periodicity\cite{0304192}, and in the limit of 
infinite periodicity
we have a single isolated D-instanton.\footnote{These solutions also 
exist for the tachyon `effective Hamiltonian' $H=\sqrt{\Pi^2 + 1/(g_s^2 
\cosh^2 T)}$.} The matrix model counterpart does
not seem to have these solutions. The resolution of this puzzle could lie
in the fact that while the quantum mechanics of a single D0-brane system
is well defined in the matrix model description, the semi-classical limit
may break down very close to the Fermi level due to the sharp cut-off on
the energy levels. Thus in the $g_s\to 0$ limit the {\it classical 
Hamiltonian}
\refb{e1} with the 
constraint \refb{e4} may not be the correct description of the system very 
close to the fermi level \refb{e2}. It is precisely at the (euclidean 
version of the) fermi level 
that a new direction
of deformation opens up for the OSFT solution. It is clearly important to
investigate this issue in detail.

\end{enumerate}

The various tests described above provide strong 
evidence that the full quantum OSFT on a single 
D0-brane is equivalent to 
the quantum theory of a single particle described by \refb{e1}, \refb{e4}. 
Indeed if the matrix model - string theory correspondence is right, 
and if the identification of the D0-brane as a single excited fermion 
is correct, then this must be the case. This, in turn, would imply that 
quantum OSFT in the continuum (1+1) dimensional string theory is an 
internally consistent theory and is capable of describing the complete 
dynamics of the D0-brane, exactly in accordance with the conjecture.

One can easily generalise this discussion to the case of multiple (say 
$n$)
D0-branes. In the matrix model 
a state of $n$ D0-branes corresponds to $n$ fermions excited from the 
fermi level to some states above the fermi level. The dynamics of such a 
system is clearly described by the quantum mechanics of
$n$ non-interacting fermions, each moving under the Hamiltonian \refb{e1} 
and satisfying the constraint
\refb{e4}. 
This is 
clearly a consistent theory by itself. 
In the continuum string theory the 
corresponding OSFT is easily 
constructed in terms of the boundary conformal field theory of $n$ 
D0-branes. The matrix model - string theory correspondence would imply 
that this quantum OSFT is exactly equivalent 
to the quantum system of $n$ fermions described above. Hence the OSFT 
describing the dynamics of $n$ D0-branes  will be 
an internally consistent quantum theory. This is again in accordance 
with our general conjecture. 
However in general ({\it e.g.} in critical string theory) even if the 
quantum OSFT on $n$ D0-branes provides a complete description of the 
system, there is no reason for this theory to be physically equivalent to
$n$ copies of the OSFT describing a single D0-brane.

We must emphasize again that in each case, the states of the D0-brane 
system, described by 
OSFT or the matrix theory hamiltonian \refb{e1}, \refb{e4}, describe only 
a subset of states in string theory. Thus we do not recover the full 
string theory by studying the OSFT, but recover a subsector of the 
theory that is a consistent quantum theory by itself. It is this lesson 
that we expect will be valid in the full critical string theory, and forms 
the basis of the conjecture stated at the beginning of this note. It is 
however instructive to ask, in the context of the two dimensional string 
theory, if there is some D0-brane system that describes the full string 
theory. To this effect we note that since a state of $n$ D0-branes 
corresponds in the matrix model to a state where $n$ fermions are excited 
above the fermi level, as we increase the number $n$ of 
D0-branes, the system is capable of describing more and more states in the 
theory. In the $n\to\infty$ limit, the states of the OSFT describe 
arbitrary excitations of multiple fermions from fermi level to any state 
above the 
fermi level. This however leaves out one important class of states, namely 
the `hole states' where we excite states from below the fermi level to the 
fermi level. If we had another set of `D0-branes' whose quantum states 
describe the hole states of the theory, then by beginning with $n$ 
usual D0-branes and $m$ `hole type' D0-branes, and taking the limit 
$m,n\to\infty$, we could represent all the states of the matrix model as 
states of the OSFT on this D0-brane system. Some proposal for the boundary 
conformal field theory describing the hole states has recently been put 
forward in \cite{0307195,0307221}, but it remains to be seen to what 
extent the dynamics 
of these `hole type' D0-branes can be described by the standard
open 
string field theory.\footnote{We note that this 
way of producing the states of string theory is somewhat different from 
the proposal of \cite{0304224,0305194}. In these papers the authors
consider taking the 
$n\to\infty$ limit 
before taking the double scaling limit of the matrix model so that the 
Fermi level is lifted by an infinite amount from its original value. 
As a result the height of the maximum of the potential above the fermi 
level changes, and we have a new string theory with a different coupling 
constant.
In 
this case the 
hole states of the new theory can be considered as particle like 
excitation in the original theory, and there is no need for exotic `hole 
type' D0-branes. In contrast we are considering the problem where the 
double scaling limit has been taken at the very beginning. Thus in this 
case the spectrum around the fermi level is continuous,
and adding any finite number $n$ of fermions does not move the fermi 
level. Thus even as we take the $n\to\infty$ limit, the Fermi level 
remains unchanged.
} Alternatively one 
might hope that by taking open string field theory on a finite number of 
space-filling non-BPS branes (or brane anti-brane systems) 
one might be able to describe all the states of the theory. However since 
there is not yet a simple description of the space-filling branes in the 
matrix model, the matrix model does not provide any insight into this 
possibility.

We should add here that the view about the hole states described 
above represents perhaps a conservative view of the situation. A 
more radical viewpoint will be  
that single hole states can also be represented as 
states of the same OSFT that describes the D0-brane. The proposal for the 
hole states outlined in \cite{0307195,0307221} involves analytic 
continuation in the parameter space of the solution describing a D0-brane 
and simultaneously changing the sign of the boundary state. It is not 
clear whether this combined operation generates a solution of the 
original 
OSFT, but it is worth examining this issue in detail. 

We would like to end with the remark that if the conjecture stated in this 
note is valid in a general string theory, then we have the possibility of 
studying different subsectors of string theory associated with different 
unstable brane systems without having to study the whole string theory at 
once. This might eventually lead to an efficient way of studying string 
theory.

{\bf Acknowledgement}: I would like to thank P.~Mukhopadhyay, L.~Rastelli,
M.~Rozali, M.~Schnabl and B.~Zwiebach for useful discussions, and
L.~Rastelli and B.~Zwiebach for a critical reading of the manuscript. I
would also like to thank the Center for Theoretical Physics at MIT, the
organisers of the Strings 2003 conference at Kyoto and the Pacific
Institute for the Mathematical Sciences at Vancouver for hospitality
during various stages of this work.

\end{document}